\documentclass[%
 reprint,
nofootinbib,
 amsmath,amssymb,
 aps,
]{revtex4-1}

\usepackage{graphicx}
\usepackage{dcolumn}
\usepackage{bm}

\begin{document}


\title[]{Reduction of Feynman Integrals in the Parametric
Representation IV: Integrals with Irregular Integration Regions}

\author{Wen Chen}
 \email{chenwenphy@gmail.com}
\affiliation{%
State Key Laboratory of Nuclear Physics and Technology, Institute of Quantum Matter, South China Normal University, Guangzhou 510006, China
}%
\affiliation{%
Guangdong Basic Research Center of Excellence for Structure and Fundamental Interactions of Matter, Guangdong Provincial Key Laboratory of Nuclear Science, Guangzhou 510006, China
}%
\date{\today}

\begin{abstract}
Parametric Feynman integrals with the regions of integration defined by some polynomials are considered in this paper. It is shown that integrals with irregular integration regions can be converted to standard parametric integrals, for which a reduction method is known. An application of this method to the analytic calculation of three-point energy correlators is presented. In principle, this method applies to more general event shapes and even jet observables.
\end{abstract}

\maketitle

\section{Introduction}
Feynman integral reduction is one of the main bottlenecks of perturbative calculations in quantum field theory. Currently, the most successful approach is the method of integration by parts (IBP)~\cite{Tkachov:1981wb,Chetyrkin:1981qh,Laporta:2000dsw}. The traditional IBP reduction is carried out in the momentum space. Alternatively, integral reduction can be carried out in the parametric representation~\cite{Regge:1968rhi,Lee:2014tja,Bitoun:2017nre,Chen:2019mqc,Chen:2019fzm,Chen:2020wsh,Artico:2023jrc,Lu:2024dsb,Wang:2024hsm}. A systematic method to reduce parametric Feynman integrals is developed by the author in a series of papers~\cite{Chen:2019mqc,Chen:2019fzm,Chen:2020wsh}, which applies to a wide class of parametric integrals. Combined with the recursive method developed in ref.~\cite{Chen:2023hmk}, this yields an automatic algorithm of Feynman integral calculation, which is implemented in the package {\bf AmpRed}~\cite{Chen:2024xwt,Chen:2025paq}.

In this paper, we consider a broader class of parametric integrals, integrals with the regions of integration defined by polynomials, known as periods in mathematics. Besides its theoretical interests, this kind of integral appears in many phenomenological applications, such as the calculations of some event shapes, including energy correlators~\cite{Dixon:2018qgp,Henn:2019gkr,Chen:2019bpb,Yan:2022cye,Chen:2023zlx,Chicherin:2024ifn,Herrmann:2024yai,Lee:2024icn,He:2025zbz,Moult:2025nhu}. Thus, a method of integral reduction for integrals with irregular integration regions is plausible.

This paper is organized as follows. We describe a method to convert integrals with irregular integration regions to standard parametric integrals in sec.~\ref{sec:ParIntIrrIntReg}. An application to the analytic calculation of the energy correlators is presented in sec.~\ref{sec:Appl}.

\section{Parametric integrals with irregular integration regions}\label{sec:ParIntIrrIntReg}

Feynman integrals (including integrals with phase-space cuts) can be converted to the following form~\cite{Chen:2019mqc,Chen:2020wsh}:
\begin{equation}\label{eq:StandParInt}
\begin{split}
&I(\lambda_0,\lambda_1,\ldots,\lambda_n)=\frac{\Gamma(-\lambda_0)}{\prod_{i=m+1}^{n+1}\Gamma(\lambda_i+1)}\\
&\quad\times\int \mathrm{d}\Pi^{(n+1)}\mathcal{F}^{\lambda_0}\prod_{i=1}^{m}x_i^{-\lambda_i-1}\prod_{i=m+1}^{n+1}x_i^{\lambda_i}~.
\end{split}
\end{equation}

\noindent Here, $\mathcal{F}(x)$ is a homogeneous polynomial of the Feynman parameters $x$ of degree $L+1$. The integration measure is $\mathrm{d}\Pi^{(n+1)}\equiv\prod_{i=1}^{n+1}\mathrm{d}x_i\delta(1-\mathcal{E}^{(1)}(x))$, with $\mathcal{E}^{(i)}(x)$ a positive definite homogeneous function of $x$ of degree $i$. The region of integration for $x_i$ is $[0,~\infty)$ if $i>m$ and $(-\infty,~\infty)$ if $i\leqslant m$. $\lambda_{n+1}\equiv-(L+1)\lambda_0-1+\sum_{i=1}^m\lambda_i-\sum_{i=m+1}^n(\lambda_i+1)$. The indices $\lambda$ may depend on some regulators. We use such a convention that $x_i^{\lambda_i}$ is understood as $(x_i-i0^+)^{\lambda_i}$ when necessary. Integrals with several factors of the form $\mathcal{F}^{\lambda_0}$ can be converted to this standard form by combining them into a single one through Mellin transforms.

In this paper, we consider integrals with irregular integration regions. That is
\begin{equation}\label{eq:BoundInt1}
J=\Gamma(-\lambda_0)\int_{\mathcal{C}}\mathrm{d}x_1\mathrm{d}x_2\cdots\mathrm{d}x_n~\mathcal{F}_0^{\lambda_0}~,
\end{equation}
Here, $\mathcal{F}_0(x)$ is not homogeneous in $x$, since we have eliminated $x_{n+1}$. The integrand may contain some factors of $x^\lambda$, but here we omit them for simplicity. The integration region $\mathcal{C}$ is defined by $\mathcal{B}_i(x)\leq0$ or $\mathcal{B}_i(x)=0$, with $\mathcal{B}_i(x)$ some polynomials in $x$. In this section, we will show that this kind of integral can be converted to the standard parametric integrals of the form in eq.~(\ref{eq:StandParInt}).

Without loss of generality, we consider the simplest case where the region of integration is defined by a single polynomial $\mathcal{B}$. Obviously, the integral in eq.~(\ref{eq:BoundInt1}) can be expressed as
\begin{equation}\label{eq:BoundInt2}
J=\Gamma(-\lambda_0)\int_{-\infty}^\infty\mathrm{d}x_1\mathrm{d}x_2\cdots\mathrm{d}x_n~\mathcal{F}_0^{\lambda_0}~\frac{w_{i}(\mathcal{B})}{2\pi}~,
\end{equation}
with $i$ being $0$ or $-1$. The $w$ function is defined by~\cite{Chen:2020wsh}
\begin{equation}\label{eq:wDef}
w_\lambda(u)\equiv e^{\frac{\lambda+1}{2}i\pi}\int_{-\infty}^{\infty}\mathrm{d}v~\frac{1}{(v+i0^+)^{\lambda+1}}e^{-ivu}~.
\end{equation}
We have
\begin{align*}
w_{-1}(u)=&~2\pi\delta(u)~,\\
w_\lambda(u)=&~2\pi\theta(u)\frac{u^\lambda}{\Gamma(\lambda+1)},~\lambda\geq 0~.\\
\end{align*}

In principle, IBP identities can directly be constructed for the integral in eq.~(\ref{eq:BoundInt2}), because the derivatives of the function $w_i$ are known~($w^{\prime}_i=w_{i-1}$). Nevertheless, this method is less systematic and there is no public code available to do this kind of reduction. Instead, in this section, we will show that the integral in eq.~(\ref{eq:BoundInt2}) can be converted to integrals of the form in eq.~(\ref{eq:StandParInt}), which can be reduced by using {\bf AmpRed}. The conversion is based on the following equation:
\begin{equation}\label{eq:Trick}
\begin{split}
    &\int_{-\infty}^\infty\mathrm{d}x~(Ax+B-i0^+)^{\lambda_0}(x-i0^+)^{\lambda}\\
    =&\left\{\begin{matrix}
        0&\quad A>0\\
        2i\pi\frac{\Gamma(-\lambda_0-\lambda-1)}{\Gamma(-\lambda_0)\Gamma(-\lambda)}B^{\lambda_0+\lambda+1}    
        (-A)^{-\lambda-1}
        &\quad A<0
    \end{matrix}\right.\\
    =&i\frac{\Gamma(-\lambda_0-\lambda-1)}{\Gamma(-\lambda_0)}B^{\lambda_0+\lambda+1}w_{-\lambda-1}(-A)~.
\end{split}
\end{equation}
This equation can easily be proven by closing the contour of integration around the branch cut. If $A>0$, all the poles of $x$ are on the same side of the real axis. Thus, the integral vanishes. If $A<0$, we close the contour of integration around the cut $Ax+B<0$. Then only the discontinuity across the cut contributes to the integral, which gives rise to a beta function.

By virtue of eq.~(\ref{eq:Trick}), we have
\begin{equation}
\begin{split}
    J=&-i\frac{1}{2\pi}\int_{-\infty}^{\infty}\mathrm{d}x_0\mathrm{d}x_1\cdots\mathrm{d}x_n~\left(\mathcal{F}_0-x_0\mathcal{B}\right)^{\lambda_0-i-1}x_0^i\\
    =&-i\frac{1}{2\pi}\int\mathrm{d}\Pi^{(n+2)}~\mathcal{F}^{\lambda_0+i}x_0^{-i-1}x_{n+1}^{\lambda_{n+1}}\\
    =&-i\frac{\Gamma(\lambda_{n+1}+1)}{2\pi}~I(\lambda_0+i,i,-1,-1,\dots)~,
\end{split}
\end{equation}
where $\lambda_{n+1}=-(L+1)\lambda_0-Li-n-1$, and $\mathcal{F}=x_{n+1}^{L+1}\left[\mathcal{F}(x/x_{n+1})-\mathcal{B}(x/x_{n+1})x_0/x_{n+1}\right]$. In the second line, we have homogenized the integrand by introducing the auxiliary variable $x_{n+1}$.

We have expressed the integral $J$ in eq.~(\ref{eq:BoundInt1}) in terms of a standard parametric integral $I$, which can be reduced by using the method developed in refs.~\cite{Chen:2019mqc,Chen:2019fzm,Chen:2020wsh}. For integrals with the integration regions defined by several polynomials, we can apply eq.~(\ref{eq:Trick}) recursively.

\section{An application}\label{sec:Appl}

As an application of the method described in the last section, we calculate the following integral:
\begin{align*}
&J(z)=\frac{(2\pi)^6}{\pi^d}\int\mathrm{d}^dk_1\mathrm{d}^dk_2~\theta(\cos\theta_{23}-z)\theta(\cos\theta_{13}-z)\\
&\quad\times\theta(\cos\theta_{12}-z)\delta(k_1^2)\delta(k_2^2)\delta(k_3^2)\frac{k_1^0k_2^0k_3^0}{(q-k_1)^2(q-k_2)^2}~,
\end{align*}
where $k_3=q-k_1-k_2$, and $\theta_{ij}$ is the angle between $\bm{k}_i$ and $\bm{k}_j$. This integral is relevant for the calculation of the three-point energy correlators~\cite{Chen:2019bpb,Yan:2022cye,Chen:2023zlx}. For simplicity, we choose $q$ to be in the time direction and set $q^2=1$. Expressing $\cos\theta_{ij}$ in terms of $k_i\cdot k_j$ and the theta functions in terms of $w_0$, eliminating all the delta functions, and integrating out the transverse components of $\bm{k}_i$, we arrive at
\begin{equation}\label{eq:E3CInt}
\begin{split}
&J(z)=\frac{\pi ^{3/2} 2^{2 \epsilon -3}}{\Gamma (1-\epsilon ) \Gamma \left(\frac{3}{2}-\epsilon \right)}\int_{-\infty}^\infty\mathrm{d}x_1\mathrm{d}x_2 \\
&\quad w_0\left(\mathcal{B}_1\right) w_0\left(\mathcal{B}_2\right)w_0\left(\mathcal{B}_3\right)\left[-x_1 x_2 \left(x_1+x_2-1\right)\right]^{-\epsilon }\\
&\quad\times x_1^{-1}x_2^{-1}\left(x_1-1\right) \left(x_2-1\right) \left(x_1+x_2\right)~,
\end{split}
\end{equation}
where $x_i=(q-k_i)^2$, and
\begin{equation*}
\begin{split}
\mathcal{B}_1=&(1-z) \left(x_1-1\right) \left(x_2-1\right)+2 \left(x_1+x_2-1\right)~,\\
\mathcal{B}_2=&-2 x_2-(1-z) \left(x_1-1\right) \left(x_1+x_2\right)~,\\
\mathcal{B}_3=&-2 x_1-(1-z) \left(x_2-1\right) \left(x_1+x_2\right)~.
\end{split}
\end{equation*}

By successive applications of eq.~(\ref{eq:Trick}), the integral in eq.~(\ref{eq:E3CInt}) can be expressed in terms of standard parametric integrals with the $\mathcal{F}$ polynomial being
\begin{equation}
\begin{split}
\mathcal{F}=&-x_1 x_2 \left(x_1+x_2-x_6\right)-x_5 \tilde{\mathcal{B}}_1-x_3\tilde{\mathcal{B}}_2-x_4 \tilde{\mathcal{B}}_3~,
\end{split}
\end{equation}
where $\tilde{\mathcal{B}}_i$ is the homogenization of $\mathcal{B}_i$. That is, $\tilde{\mathcal{B}}_i(x)\equiv x_6^2\mathcal{B}(x/x_6)$. Explicitly, we have
\begin{equation*}
\begin{split}
J=&\frac{i \pi ^{3/2} 2^{2 \epsilon -3} \Gamma (3 \epsilon -3)}{\Gamma (1-\epsilon ) \Gamma \left(\frac{3}{2}-\epsilon \right) \Gamma (\epsilon )}\\
&\times\Big\{(\mathcal{D}_1+\mathcal{D}_2-6 \epsilon +6) I(-\epsilon ,-1,-1,0,0,0)\\
&+3 (\epsilon -1) \big[(-\mathcal{D}_1+3 \epsilon -2) I(-\epsilon ,-1,0,0,0,0)\\
&+(-\mathcal{D}_2+3 \epsilon -2) I(-\epsilon ,0,-1,0,0,0)\big]\Big\}~,
\end{split}
\end{equation*}
where $\epsilon=(4-d)/2$, and $\mathcal{D}_i$ is an operator defined by
\begin{equation*}
\mathcal{D}_i~I(\lambda_0,\lambda_1,\dots,\lambda_i,\dots)=I(\lambda_0,\lambda_1,\dots,\lambda_i-1,\dots)~.
\end{equation*}

The obtained integrals can further be reduced by using {\bf AmpRed}. We get
\begin{equation*}
\begin{split}
J=&-\frac{i \pi \Gamma (3 \epsilon -2)}{6 (3 \epsilon -1) \Gamma (2-2 \epsilon ) \Gamma (\epsilon )}\\
&\times\Big\{\left[2 (3 \epsilon -1)^2-2 z (\epsilon -2) (3 \epsilon -1)\right]I_2 \\
&+ \left[2 z (\epsilon -2)-10 (2 \epsilon -1)\right]I_3\\
&- \epsilon  (3 \epsilon -1) (4 \epsilon -3)I_4\Big\}~,
\end{split}
\end{equation*}
where the master integrals $I_i$ are\footnote{Here we choose the master integrals manually to simplify the result. We have reduced the number of master integrals by using the symmetry between $k_1$, $k_2$, and $k_3$. Because we choose $(q-k_1)^2$ and $(q-k_2)^2$ as the independent variables, the explicit symmetry is broken, and thus can not be automatically detected by {\bf AmpRed}.}
\begin{align*}
I_1 = & I(-1 - \epsilon,-1,-1,-1,-1,0)~,\\
I_2 = & I(-1 - \epsilon,-1,-1,0,-1,0)~,\\
I_3 = & I(-1 - \epsilon,-1,-2,0,-1,0)~,\\
I_4 = & I(-1 - \epsilon,-1,-1,0,0,0)~.
\end{align*}
The differential equations of the master integrals are
\begin{equation*}
\frac{\mathrm{d}\bm{I}}{\mathrm{d}z}=M\cdot \bm{I}~,
\end{equation*}
where
\begin{equation*}
\begin{split}
&M=\\
&\begin{pmatrix}
 -\frac{(2 z+1) (z-2 \epsilon +1)}{(z-1) z (z+1)} & 0 & 0 & 0 \\
 -\frac{4 (z-1) (\epsilon -1)}{z (z+1) (3 \epsilon -1)} & -\frac{2 z \
\epsilon }{(z-1) (z+1)} & \frac{2 \epsilon -1}{(z+1) (3 \epsilon -1)} \
& 0 \\
 0 & -\frac{(2 \epsilon -1) (3 \epsilon -1)}{z-1} & \frac{2 (z \
\epsilon -z+2 \epsilon -1)}{(z-1) (z+1)} & 0 \\
 \frac{8 (z-1) (\epsilon -1)}{3 z \epsilon  (3 \epsilon -1)} & \
\frac{4 (3 z \epsilon -z+5 \epsilon -1)}{3 (z-1) (z+1) \epsilon } & -\
\frac{4 (2 \epsilon -1)}{3 (z+1) \epsilon  (3 \epsilon -1)} & 0 
\end{pmatrix}~.
\end{split}
\end{equation*}
The differential-equation system can easily be transformed into the canonical form by using {\tt Libra}~\cite{Lee:2014ioa,Lee:2020zfb}. We choose the boundary of the differential equations to be at $z=-1/2$, which corresponds to $\max{\theta_{ij}}=2\pi/3$. $I_1$ can be trivially calculated by converting it back to the representation of eq.~(\ref{eq:E3CInt}) and eliminating the integration variables using the two delta functions. All the other master integrals vanish at $z=-1/2$, because there is no available phase space. Hence, the boundary conditions of the differential equations are trivial. Here we omit the details of the remaining calculations. Finally, we get
\begin{equation*}
\begin{split}
&J(z)=\frac{216 \pi ^4 \epsilon  \Gamma (\epsilon +1) \Gamma (3 \epsilon -3)}{(z-1)^2 \Gamma (2-2 \epsilon )}\\
&\quad\times\left[(2z+1)(2z-5)+6 \log (2 (z+1))+\mathcal{O}(\epsilon)\right]~.
\end{split}
\end{equation*}

\section{Summary and discussion}
Parametric Feynman integrals with integration regions defined by polynomials are considered in this paper. It is shown that integrals with irregular integration regions can be converted to standard parametric integrals and thus can be reduced by using the method developed in refs.~\cite{Chen:2019mqc,Chen:2019fzm,Chen:2020wsh}. As an application of this method, we analytically calculate an integral related to the three-point energy correlators.

Notice that for most event shapes, the phase-space constraints can be expressed in terms of polynomial equations of Lorentz scalars, which are linear combinations of the Baikov variables in the Baikov representation~\cite{Cutkosky:1960sp,Baikov:1996iu}. Integrals in the Baikov representation are of the structure of eq.~(\ref{eq:StandParInt})~\footnote{It can be shown that the Baikov representation is identical to the dual representation of the parametric representation~\cite{Chen:2023eqx}.}. Thus, in principle, the method developed in this paper applies to more general event shapes and even jet observables.

\begin{acknowledgments}
This work is supported by National Natural Science Foundation of China~(NSFC) under Grant No. 12405095 and Guangdong Major Project of Basic and Applied Basic Research~(No. 2020B0301030008).
\end{acknowledgments}

\bibliography{refs}

\begin{thebibliography}{30}%
\makeatletter
\providecommand \@ifxundefined [1]{%
 \@ifx{#1\undefined}
}%
\providecommand \@ifnum [1]{%
 \ifnum #1\expandafter \@firstoftwo
 \else \expandafter \@secondoftwo
 \fi
}%
\providecommand \@ifx [1]{%
 \ifx #1\expandafter \@firstoftwo
 \else \expandafter \@secondoftwo
 \fi
}%
\providecommand \natexlab [1]{#1}%
\providecommand \enquote  [1]{``#1''}%
\providecommand \bibnamefont  [1]{#1}%
\providecommand \bibfnamefont [1]{#1}%
\providecommand \citenamefont [1]{#1}%
\providecommand \href@noop [0]{\@secondoftwo}%
\providecommand \href [0]{\begingroup \@sanitize@url \@href}%
\providecommand \@href[1]{\@@startlink{#1}\@@href}%
\providecommand \@@href[1]{\endgroup#1\@@endlink}%
\providecommand \@sanitize@url [0]{\catcode `\\12\catcode `\$12\catcode
  `\&12\catcode `\#12\catcode `\^12\catcode `\_12\catcode `\%12\relax}%
\providecommand \@@startlink[1]{}%
\providecommand \@@endlink[0]{}%
\providecommand \url  [0]{\begingroup\@sanitize@url \@url }%
\providecommand \@url [1]{\endgroup\@href {#1}{\urlprefix }}%
\providecommand \urlprefix  [0]{URL }%
\providecommand \Eprint [0]{\href }%
\providecommand \doibase [0]{http://dx.doi.org/}%
\providecommand \selectlanguage [0]{\@gobble}%
\providecommand \bibinfo  [0]{\@secondoftwo}%
\providecommand \bibfield  [0]{\@secondoftwo}%
\providecommand \translation [1]{[#1]}%
\providecommand \BibitemOpen [0]{}%
\providecommand \bibitemStop [0]{}%
\providecommand \bibitemNoStop [0]{.\EOS\space}%
\providecommand \EOS [0]{\spacefactor3000\relax}%
\providecommand \BibitemShut  [1]{\csname bibitem#1\endcsname}%
\let\auto@bib@innerbib\@empty
\bibitem [{\citenamefont {Tkachov}(1981)}]{Tkachov:1981wb}%
  \BibitemOpen
  \bibfield  {author} {\bibinfo {author} {\bibfnamefont {F.~V.}\ \bibnamefont
  {Tkachov}},\ }\href {\doibase 10.1016/0370-2693(81)90288-4} {\bibfield
  {journal} {\bibinfo  {journal} {Phys. Lett. B}\ }\textbf {\bibinfo {volume}
  {100}},\ \bibinfo {pages} {65} (\bibinfo {year} {1981})}\BibitemShut
  {NoStop}%
\bibitem [{\citenamefont {Chetyrkin}\ and\ \citenamefont
  {Tkachov}(1981)}]{Chetyrkin:1981qh}%
  \BibitemOpen
  \bibfield  {author} {\bibinfo {author} {\bibfnamefont {K.~G.}\ \bibnamefont
  {Chetyrkin}}\ and\ \bibinfo {author} {\bibfnamefont {F.~V.}\ \bibnamefont
  {Tkachov}},\ }\href {\doibase 10.1016/0550-3213(81)90199-1} {\bibfield
  {journal} {\bibinfo  {journal} {Nucl. Phys. B}\ }\textbf {\bibinfo {volume}
  {192}},\ \bibinfo {pages} {159} (\bibinfo {year} {1981})}\BibitemShut
  {NoStop}%
\bibitem [{\citenamefont {Laporta}(2000)}]{Laporta:2000dsw}%
  \BibitemOpen
  \bibfield  {author} {\bibinfo {author} {\bibfnamefont {S.}~\bibnamefont
  {Laporta}},\ }\href {\doibase 10.1142/S0217751X00002159} {\bibfield
  {journal} {\bibinfo  {journal} {Int. J. Mod. Phys. A}\ }\textbf {\bibinfo
  {volume} {15}},\ \bibinfo {pages} {5087} (\bibinfo {year} {2000})},\ \Eprint
  {http://arxiv.org/abs/hep-ph/0102033} {arXiv:hep-ph/0102033} \BibitemShut
  {NoStop}%
\bibitem [{\citenamefont {Regge}(1968)}]{Regge:1968rhi}%
  \BibitemOpen
  \bibfield  {author} {\bibinfo {author} {\bibfnamefont {T.}~\bibnamefont
  {Regge}},\ }in\ \href@noop {} {\emph {\bibinfo {booktitle} {{Battelle
  Rencontres}}}}\ (\bibinfo {year} {1968})\ pp.\ \bibinfo {pages}
  {433--458}\BibitemShut {NoStop}%
\bibitem [{\citenamefont {Lee}(2014)}]{Lee:2014tja}%
  \BibitemOpen
  \bibfield  {author} {\bibinfo {author} {\bibfnamefont {R.~N.}\ \bibnamefont
  {Lee}},\ }in\ \href@noop {} {\emph {\bibinfo {booktitle} {{49th Rencontres de
  Moriond on QCD and High Energy Interactions}}}}\ (\bibinfo {year} {2014})\
  pp.\ \bibinfo {pages} {297--300},\ \Eprint {http://arxiv.org/abs/1405.5616}
  {arXiv:1405.5616 [hep-ph]} \BibitemShut {NoStop}%
\bibitem [{\citenamefont {Bitoun}\ \emph {et~al.}(2019)\citenamefont {Bitoun},
  \citenamefont {Bogner}, \citenamefont {Klausen},\ and\ \citenamefont
  {Panzer}}]{Bitoun:2017nre}%
  \BibitemOpen
  \bibfield  {author} {\bibinfo {author} {\bibfnamefont {T.}~\bibnamefont
  {Bitoun}}, \bibinfo {author} {\bibfnamefont {C.}~\bibnamefont {Bogner}},
  \bibinfo {author} {\bibfnamefont {R.~P.}\ \bibnamefont {Klausen}}, \ and\
  \bibinfo {author} {\bibfnamefont {E.}~\bibnamefont {Panzer}},\ }\href
  {\doibase 10.1007/s11005-018-1114-8} {\bibfield  {journal} {\bibinfo
  {journal} {Lett. Math. Phys.}\ }\textbf {\bibinfo {volume} {109}},\ \bibinfo
  {pages} {497} (\bibinfo {year} {2019})},\ \Eprint
  {http://arxiv.org/abs/1712.09215} {arXiv:1712.09215 [hep-th]} \BibitemShut
  {NoStop}%
\bibitem [{\citenamefont {Chen}(2020{\natexlab{a}})}]{Chen:2019mqc}%
  \BibitemOpen
  \bibfield  {author} {\bibinfo {author} {\bibfnamefont {W.}~\bibnamefont
  {Chen}},\ }\href {\doibase 10.1007/JHEP02(2020)115} {\bibfield  {journal}
  {\bibinfo  {journal} {JHEP}\ }\textbf {\bibinfo {volume} {02}},\ \bibinfo
  {pages} {115} (\bibinfo {year} {2020}{\natexlab{a}})},\ \Eprint
  {http://arxiv.org/abs/1902.10387} {arXiv:1902.10387 [hep-ph]} \BibitemShut
  {NoStop}%
\bibitem [{\citenamefont {Chen}(2021)}]{Chen:2019fzm}%
  \BibitemOpen
  \bibfield  {author} {\bibinfo {author} {\bibfnamefont {W.}~\bibnamefont
  {Chen}},\ }\href {\doibase 10.1140/epjc/s10052-021-09036-5} {\bibfield
  {journal} {\bibinfo  {journal} {Eur. Phys. J. C}\ }\textbf {\bibinfo {volume}
  {81}},\ \bibinfo {pages} {244} (\bibinfo {year} {2021})},\ \Eprint
  {http://arxiv.org/abs/1912.08606} {arXiv:1912.08606 [hep-ph]} \BibitemShut
  {NoStop}%
\bibitem [{\citenamefont {Chen}(2020{\natexlab{b}})}]{Chen:2020wsh}%
  \BibitemOpen
  \bibfield  {author} {\bibinfo {author} {\bibfnamefont {W.}~\bibnamefont
  {Chen}},\ }\href {\doibase 10.1140/epjc/s10052-020-08757-3} {\bibfield
  {journal} {\bibinfo  {journal} {Eur. Phys. J. C}\ }\textbf {\bibinfo {volume}
  {80}},\ \bibinfo {pages} {1173} (\bibinfo {year} {2020}{\natexlab{b}})},\
  \Eprint {http://arxiv.org/abs/2007.00507} {arXiv:2007.00507 [hep-ph]}
  \BibitemShut {NoStop}%
\bibitem [{\citenamefont {Artico}\ and\ \citenamefont
  {Magnea}(2024)}]{Artico:2023jrc}%
  \BibitemOpen
  \bibfield  {author} {\bibinfo {author} {\bibfnamefont {D.}~\bibnamefont
  {Artico}}\ and\ \bibinfo {author} {\bibfnamefont {L.}~\bibnamefont
  {Magnea}},\ }\href {\doibase 10.1007/JHEP03(2024)096} {\bibfield  {journal}
  {\bibinfo  {journal} {JHEP}\ }\textbf {\bibinfo {volume} {03}},\ \bibinfo
  {pages} {096} (\bibinfo {year} {2024})},\ \Eprint
  {http://arxiv.org/abs/2310.03939} {arXiv:2310.03939 [hep-ph]} \BibitemShut
  {NoStop}%
\bibitem [{\citenamefont {Lu}\ \emph {et~al.}(2024)\citenamefont {Lu},
  \citenamefont {Wang},\ and\ \citenamefont {Yang}}]{Lu:2024dsb}%
  \BibitemOpen
  \bibfield  {author} {\bibinfo {author} {\bibfnamefont {M.}~\bibnamefont
  {Lu}}, \bibinfo {author} {\bibfnamefont {Z.}~\bibnamefont {Wang}}, \ and\
  \bibinfo {author} {\bibfnamefont {L.~L.}\ \bibnamefont {Yang}},\ }\href@noop
  {} {\  (\bibinfo {year} {2024})},\ \Eprint {http://arxiv.org/abs/2411.05226}
  {arXiv:2411.05226 [hep-th]} \BibitemShut {NoStop}%
\bibitem [{\citenamefont {Wang}\ and\ \citenamefont
  {Yang}(2024)}]{Wang:2024hsm}%
  \BibitemOpen
  \bibfield  {author} {\bibinfo {author} {\bibfnamefont {Z.}~\bibnamefont
  {Wang}}\ and\ \bibinfo {author} {\bibfnamefont {L.~L.}\ \bibnamefont
  {Yang}},\ }\href@noop {} {\  (\bibinfo {year} {2024})},\ \Eprint
  {http://arxiv.org/abs/2412.15962} {arXiv:2412.15962 [hep-th]} \BibitemShut
  {NoStop}%
\bibitem [{\citenamefont {Chen}\ \emph
  {et~al.}(2024{\natexlab{a}})\citenamefont {Chen}, \citenamefont {Luo},
  \citenamefont {Yang},\ and\ \citenamefont {Zhu}}]{Chen:2023hmk}%
  \BibitemOpen
  \bibfield  {author} {\bibinfo {author} {\bibfnamefont {W.}~\bibnamefont
  {Chen}}, \bibinfo {author} {\bibfnamefont {M.-x.}\ \bibnamefont {Luo}},
  \bibinfo {author} {\bibfnamefont {T.-Z.}\ \bibnamefont {Yang}}, \ and\
  \bibinfo {author} {\bibfnamefont {H.~X.}\ \bibnamefont {Zhu}},\ }\href
  {\doibase 10.1007/JHEP01(2024)131} {\bibfield  {journal} {\bibinfo  {journal}
  {JHEP}\ }\textbf {\bibinfo {volume} {01}},\ \bibinfo {pages} {131} (\bibinfo
  {year} {2024}{\natexlab{a}})},\ \Eprint {http://arxiv.org/abs/2309.03832}
  {arXiv:2309.03832 [hep-ph]} \BibitemShut {NoStop}%
\bibitem [{\citenamefont {Chen}(2025{\natexlab{a}})}]{Chen:2024xwt}%
  \BibitemOpen
  \bibfield  {author} {\bibinfo {author} {\bibfnamefont {W.}~\bibnamefont
  {Chen}},\ }\href {\doibase 10.1016/j.cpc.2025.109607} {\bibfield  {journal}
  {\bibinfo  {journal} {Comput. Phys. Commun.}\ }\textbf {\bibinfo {volume}
  {312}},\ \bibinfo {pages} {109607} (\bibinfo {year} {2025}{\natexlab{a}})},\
  \Eprint {http://arxiv.org/abs/2408.06426} {arXiv:2408.06426 [hep-ph]}
  \BibitemShut {NoStop}%
\bibitem [{\citenamefont {Chen}(2025{\natexlab{b}})}]{Chen:2025paq}%
  \BibitemOpen
  \bibfield  {author} {\bibinfo {author} {\bibfnamefont {W.}~\bibnamefont
  {Chen}},\ }\href@noop {} {\  (\bibinfo {year} {2025}{\natexlab{b}})},\
  \Eprint {http://arxiv.org/abs/2505.13540} {arXiv:2505.13540 [hep-ph]}
  \BibitemShut {NoStop}%
\bibitem [{\citenamefont {Dixon}\ \emph {et~al.}(2018)\citenamefont {Dixon},
  \citenamefont {Luo}, \citenamefont {Shtabovenko}, \citenamefont {Yang},\ and\
  \citenamefont {Zhu}}]{Dixon:2018qgp}%
  \BibitemOpen
  \bibfield  {author} {\bibinfo {author} {\bibfnamefont {L.~J.}\ \bibnamefont
  {Dixon}}, \bibinfo {author} {\bibfnamefont {M.-X.}\ \bibnamefont {Luo}},
  \bibinfo {author} {\bibfnamefont {V.}~\bibnamefont {Shtabovenko}}, \bibinfo
  {author} {\bibfnamefont {T.-Z.}\ \bibnamefont {Yang}}, \ and\ \bibinfo
  {author} {\bibfnamefont {H.~X.}\ \bibnamefont {Zhu}},\ }\href {\doibase
  10.1103/PhysRevLett.120.102001} {\bibfield  {journal} {\bibinfo  {journal}
  {Phys. Rev. Lett.}\ }\textbf {\bibinfo {volume} {120}},\ \bibinfo {pages}
  {102001} (\bibinfo {year} {2018})},\ \Eprint
  {http://arxiv.org/abs/1801.03219} {arXiv:1801.03219 [hep-ph]} \BibitemShut
  {NoStop}%
\bibitem [{\citenamefont {Henn}\ \emph {et~al.}(2019)\citenamefont {Henn},
  \citenamefont {Sokatchev}, \citenamefont {Yan},\ and\ \citenamefont
  {Zhiboedov}}]{Henn:2019gkr}%
  \BibitemOpen
  \bibfield  {author} {\bibinfo {author} {\bibfnamefont {J.~M.}\ \bibnamefont
  {Henn}}, \bibinfo {author} {\bibfnamefont {E.}~\bibnamefont {Sokatchev}},
  \bibinfo {author} {\bibfnamefont {K.}~\bibnamefont {Yan}}, \ and\ \bibinfo
  {author} {\bibfnamefont {A.}~\bibnamefont {Zhiboedov}},\ }\href {\doibase
  10.1103/PhysRevD.100.036010} {\bibfield  {journal} {\bibinfo  {journal}
  {Phys. Rev. D}\ }\textbf {\bibinfo {volume} {100}},\ \bibinfo {pages}
  {036010} (\bibinfo {year} {2019})},\ \Eprint
  {http://arxiv.org/abs/1903.05314} {arXiv:1903.05314 [hep-th]} \BibitemShut
  {NoStop}%
\bibitem [{\citenamefont {Chen}\ \emph {et~al.}(2020)\citenamefont {Chen},
  \citenamefont {Luo}, \citenamefont {Moult}, \citenamefont {Yang},
  \citenamefont {Zhang},\ and\ \citenamefont {Zhu}}]{Chen:2019bpb}%
  \BibitemOpen
  \bibfield  {author} {\bibinfo {author} {\bibfnamefont {H.}~\bibnamefont
  {Chen}}, \bibinfo {author} {\bibfnamefont {M.-X.}\ \bibnamefont {Luo}},
  \bibinfo {author} {\bibfnamefont {I.}~\bibnamefont {Moult}}, \bibinfo
  {author} {\bibfnamefont {T.-Z.}\ \bibnamefont {Yang}}, \bibinfo {author}
  {\bibfnamefont {X.}~\bibnamefont {Zhang}}, \ and\ \bibinfo {author}
  {\bibfnamefont {H.~X.}\ \bibnamefont {Zhu}},\ }\href {\doibase
  10.1007/JHEP08(2020)028} {\bibfield  {journal} {\bibinfo  {journal} {JHEP}\
  }\textbf {\bibinfo {volume} {08}},\ \bibinfo {pages} {028} (\bibinfo {year}
  {2020})},\ \Eprint {http://arxiv.org/abs/1912.11050} {arXiv:1912.11050
  [hep-ph]} \BibitemShut {NoStop}%
\bibitem [{\citenamefont {Yan}\ and\ \citenamefont
  {Zhang}(2022)}]{Yan:2022cye}%
  \BibitemOpen
  \bibfield  {author} {\bibinfo {author} {\bibfnamefont {K.}~\bibnamefont
  {Yan}}\ and\ \bibinfo {author} {\bibfnamefont {X.}~\bibnamefont {Zhang}},\
  }\href {\doibase 10.1103/PhysRevLett.129.021602} {\bibfield  {journal}
  {\bibinfo  {journal} {Phys. Rev. Lett.}\ }\textbf {\bibinfo {volume} {129}},\
  \bibinfo {pages} {021602} (\bibinfo {year} {2022})},\ \Eprint
  {http://arxiv.org/abs/2203.04349} {arXiv:2203.04349 [hep-th]} \BibitemShut
  {NoStop}%
\bibitem [{\citenamefont {Chen}\ \emph
  {et~al.}(2024{\natexlab{b}})\citenamefont {Chen}, \citenamefont {Gao},
  \citenamefont {Li}, \citenamefont {Xu}, \citenamefont {Zhang},\ and\
  \citenamefont {Zhu}}]{Chen:2023zlx}%
  \BibitemOpen
  \bibfield  {author} {\bibinfo {author} {\bibfnamefont {W.}~\bibnamefont
  {Chen}}, \bibinfo {author} {\bibfnamefont {J.}~\bibnamefont {Gao}}, \bibinfo
  {author} {\bibfnamefont {Y.}~\bibnamefont {Li}}, \bibinfo {author}
  {\bibfnamefont {Z.}~\bibnamefont {Xu}}, \bibinfo {author} {\bibfnamefont
  {X.}~\bibnamefont {Zhang}}, \ and\ \bibinfo {author} {\bibfnamefont {H.~X.}\
  \bibnamefont {Zhu}},\ }\href {\doibase 10.1007/JHEP05(2024)043} {\bibfield
  {journal} {\bibinfo  {journal} {JHEP}\ }\textbf {\bibinfo {volume} {05}},\
  \bibinfo {pages} {043} (\bibinfo {year} {2024}{\natexlab{b}})},\ \Eprint
  {http://arxiv.org/abs/2307.07510} {arXiv:2307.07510 [hep-ph]} \BibitemShut
  {NoStop}%
\bibitem [{\citenamefont {Chicherin}\ \emph {et~al.}(2024)\citenamefont
  {Chicherin}, \citenamefont {Moult}, \citenamefont {Sokatchev}, \citenamefont
  {Yan},\ and\ \citenamefont {Zhu}}]{Chicherin:2024ifn}%
  \BibitemOpen
  \bibfield  {author} {\bibinfo {author} {\bibfnamefont {D.}~\bibnamefont
  {Chicherin}}, \bibinfo {author} {\bibfnamefont {I.}~\bibnamefont {Moult}},
  \bibinfo {author} {\bibfnamefont {E.}~\bibnamefont {Sokatchev}}, \bibinfo
  {author} {\bibfnamefont {K.}~\bibnamefont {Yan}}, \ and\ \bibinfo {author}
  {\bibfnamefont {Y.}~\bibnamefont {Zhu}},\ }\href {\doibase
  10.1103/PhysRevD.110.L091901} {\bibfield  {journal} {\bibinfo  {journal}
  {Phys. Rev. D}\ }\textbf {\bibinfo {volume} {110}},\ \bibinfo {pages}
  {L091901} (\bibinfo {year} {2024})},\ \Eprint
  {http://arxiv.org/abs/2401.06463} {arXiv:2401.06463 [hep-th]} \BibitemShut
  {NoStop}%
\bibitem [{\citenamefont {Herrmann}\ \emph {et~al.}(2024)\citenamefont
  {Herrmann}, \citenamefont {Kologlu},\ and\ \citenamefont
  {Moult}}]{Herrmann:2024yai}%
  \BibitemOpen
  \bibfield  {author} {\bibinfo {author} {\bibfnamefont {E.}~\bibnamefont
  {Herrmann}}, \bibinfo {author} {\bibfnamefont {M.}~\bibnamefont {Kologlu}}, \
  and\ \bibinfo {author} {\bibfnamefont {I.}~\bibnamefont {Moult}},\
  }\href@noop {} {\  (\bibinfo {year} {2024})},\ \Eprint
  {http://arxiv.org/abs/2412.05384} {arXiv:2412.05384 [hep-th]} \BibitemShut
  {NoStop}%
\bibitem [{\citenamefont {Lee}\ \emph {et~al.}(2025)\citenamefont {Lee},
  \citenamefont {Moult},\ and\ \citenamefont {Zhang}}]{Lee:2024icn}%
  \BibitemOpen
  \bibfield  {author} {\bibinfo {author} {\bibfnamefont {K.}~\bibnamefont
  {Lee}}, \bibinfo {author} {\bibfnamefont {I.}~\bibnamefont {Moult}}, \ and\
  \bibinfo {author} {\bibfnamefont {X.}~\bibnamefont {Zhang}},\ }\href
  {\doibase 10.1007/JHEP05(2025)129} {\bibfield  {journal} {\bibinfo  {journal}
  {JHEP}\ }\textbf {\bibinfo {volume} {05}},\ \bibinfo {pages} {129} (\bibinfo
  {year} {2025})},\ \Eprint {http://arxiv.org/abs/2409.19045} {arXiv:2409.19045
  [hep-ph]} \BibitemShut {NoStop}%
\bibitem [{\citenamefont {He}\ \emph {et~al.}(2025)\citenamefont {He},
  \citenamefont {Li}, \citenamefont {Lin}, \citenamefont {Liu},\ and\
  \citenamefont {Yan}}]{He:2025zbz}%
  \BibitemOpen
  \bibfield  {author} {\bibinfo {author} {\bibfnamefont {S.}~\bibnamefont
  {He}}, \bibinfo {author} {\bibfnamefont {X.}~\bibnamefont {Li}}, \bibinfo
  {author} {\bibfnamefont {J.}~\bibnamefont {Lin}}, \bibinfo {author}
  {\bibfnamefont {J.}~\bibnamefont {Liu}}, \ and\ \bibinfo {author}
  {\bibfnamefont {K.}~\bibnamefont {Yan}},\ }\href@noop {} {\  (\bibinfo {year}
  {2025})},\ \Eprint {http://arxiv.org/abs/2506.07796} {arXiv:2506.07796
  [hep-th]} \BibitemShut {NoStop}%
\bibitem [{\citenamefont {Moult}\ and\ \citenamefont
  {Zhu}(2025)}]{Moult:2025nhu}%
  \BibitemOpen
  \bibfield  {author} {\bibinfo {author} {\bibfnamefont {I.}~\bibnamefont
  {Moult}}\ and\ \bibinfo {author} {\bibfnamefont {H.~X.}\ \bibnamefont
  {Zhu}},\ }\href@noop {} {\  (\bibinfo {year} {2025})},\ \Eprint
  {http://arxiv.org/abs/2506.09119} {arXiv:2506.09119 [hep-ph]} \BibitemShut
  {NoStop}%
\bibitem [{\citenamefont {Lee}(2015)}]{Lee:2014ioa}%
  \BibitemOpen
  \bibfield  {author} {\bibinfo {author} {\bibfnamefont {R.~N.}\ \bibnamefont
  {Lee}},\ }\href {\doibase 10.1007/JHEP04(2015)108} {\bibfield  {journal}
  {\bibinfo  {journal} {JHEP}\ }\textbf {\bibinfo {volume} {04}},\ \bibinfo
  {pages} {108} (\bibinfo {year} {2015})},\ \Eprint
  {http://arxiv.org/abs/1411.0911} {arXiv:1411.0911 [hep-ph]} \BibitemShut
  {NoStop}%
\bibitem [{\citenamefont {Lee}(2021)}]{Lee:2020zfb}%
  \BibitemOpen
  \bibfield  {author} {\bibinfo {author} {\bibfnamefont {R.~N.}\ \bibnamefont
  {Lee}},\ }\href {\doibase 10.1016/j.cpc.2021.108058} {\bibfield  {journal}
  {\bibinfo  {journal} {Comput. Phys. Commun.}\ }\textbf {\bibinfo {volume}
  {267}},\ \bibinfo {pages} {108058} (\bibinfo {year} {2021})},\ \Eprint
  {http://arxiv.org/abs/2012.00279} {arXiv:2012.00279 [hep-ph]} \BibitemShut
  {NoStop}%
\bibitem [{\citenamefont {Cutkosky}(1960)}]{Cutkosky:1960sp}%
  \BibitemOpen
  \bibfield  {author} {\bibinfo {author} {\bibfnamefont {R.~E.}\ \bibnamefont
  {Cutkosky}},\ }\href {\doibase 10.1063/1.1703676} {\bibfield  {journal}
  {\bibinfo  {journal} {J. Math. Phys.}\ }\textbf {\bibinfo {volume} {1}},\
  \bibinfo {pages} {429} (\bibinfo {year} {1960})}\BibitemShut {NoStop}%
\bibitem [{\citenamefont {Baikov}(1997)}]{Baikov:1996iu}%
  \BibitemOpen
  \bibfield  {author} {\bibinfo {author} {\bibfnamefont {P.~A.}\ \bibnamefont
  {Baikov}},\ }\href {\doibase 10.1016/S0168-9002(97)00126-5} {\bibfield
  {journal} {\bibinfo  {journal} {Nucl. Instrum. Meth. A}\ }\textbf {\bibinfo
  {volume} {389}},\ \bibinfo {pages} {347} (\bibinfo {year} {1997})},\ \Eprint
  {http://arxiv.org/abs/hep-ph/9611449} {arXiv:hep-ph/9611449} \BibitemShut
  {NoStop}%
\bibitem [{\citenamefont {Chen}(2025{\natexlab{c}})}]{Chen:2023eqx}%
  \BibitemOpen
  \bibfield  {author} {\bibinfo {author} {\bibfnamefont {W.}~\bibnamefont
  {Chen}},\ }\href {\doibase 10.1016/j.physletb.2025.139340} {\bibfield
  {journal} {\bibinfo  {journal} {Phys. Lett. B}\ }\textbf {\bibinfo {volume}
  {862}},\ \bibinfo {pages} {139340} (\bibinfo {year} {2025}{\natexlab{c}})},\
  \Eprint {http://arxiv.org/abs/2303.12427} {arXiv:2303.12427 [hep-ph]}
  \BibitemShut {NoStop}%
\end{thebibliography}%

\end{document}